\documentclass{ws-procs975x65}

\usepackage[utf8x]{inputenc}
\usepackage{graphicx}
\usepackage{latexsym}
\usepackage{amsmath}

\usepackage{pdfsync}

\newcommand{\Tr}{\textrm{Tr}\,}

\newcommand{\ev}[1]{\langle #1 \rangle}

\newcommand{\csw}{c_{\rm{sw}}}

\newcommand{\be}{\begin{equation}}
\newcommand{\ee}{\end{equation}}
\newcommand{\bea}{\begin{eqnarray}} 
\newcommand{\eea}{\end{eqnarray}}

\newcommand{\bmp}{\noindent\begin{minipage}{16cm}}
\newcommand{\emp}{\end{minipage}\vskip 7mm} 
\def\lsim{\mathrel{\raise.3ex\hbox{$<$\kern-.75em\lower1ex\hbox{$\sim$}}}}
\def\gsim{\mathrel{\raise.3ex\hbox{$>$\kern-.75em\lower1ex\hbox{$\sim$}}}}

\newcommand{\intron}[1]{}

\begin{document}

\title{Running coupling in SU(2) with adjoint fermions}
\author{Jarno Rantaharju\footnote{jarno.rantaharju@riken.jp}}
\address{Riken Advanced Institute for Computational Science,\\
 7-1-26, Minatojima-minami-machi, Chuo-ku, Kobe, Hyogo, Japan}
\author{Kari Rummukainen\footnote{kari.rummukainen@helsinki.fi}}
\address{
 Department of Physics and Helsinki Institute of Physics,\\
 P.O.Box 64, FI-00014 University of Helsinki, Finland}
\author{Kimmo Tuominen\footnote{kimmo.i.tuominen@jyu.fi}}
\address{
Department of Physics, P.O.Box 35 (YFL), 
        \\ FI-40014 University of Jyv\"askyl\"a, Finland, 
        \\ and 
  	    \\ Helsinki Institute of Physics, P.O.~Box 64, 
  	    \\ FI-00014 University of Helsinki, Finland.}
\abstract {%
  We present a measurement of the Schr\"odinger Functional running coupling in SU(2) lattice gauge theory with adjoint fermions. We use HEX smearing and clover improvement to reduce the discretization effects. We obtain a robust continuum limit for the step scaling, which confirms the existence of a non-trivial fixed point.
}


\bodymatter

\section{The Model}

We study the running of the Schr\"odinger Functional coupling in the SU(2) lattice field theory with 2 fermions in the adjoint representation. The model, dubbed Minimal Walking Technicolor, has been studied recently as a possible candidate for a walking technicolor theory and as a part of the ongoing mapping of the conformal window on the lattice \cite{ Bursa:2011ru,Hietanen:2008mr,Bursa:2009tj,Hietanen:2009az,DeGrand:2011qd}. We define the lattice model by the action
\begin{equation}
 S = S_G + S_F,
\end{equation}
where $S_F$ is the clover improved Wilson fermion action with smeared gauge links and $S_G$ is a partially smeared version of the Wilson plaquette action. We use hypercubic stout smearing, or HEX smearing,\cite{Durr:2010aw} 
to reduce the discretization errors and allow simulations at larger coupling.
The gauge action is defined as
\begin{align}
  S_G
  &= \beta_L \sum_{x;\mu<\nu} (1-c_g) {\mathcal L}_{x,\mu}(U) + c_g {\mathcal L}_{x,\mu}(V) \\
  {\mathcal L}_{x,\mu}(U) &= \left (1 - \frac12 \Tr [U_\mu(x) U_\nu(x+a\hat\mu) 
  U^\dagger_\mu(x+a\hat\nu) U^\dagger_\nu(x) ] \right),\nonumber
\end{align}
where $V$ is the smeared gauge field and we choose  $c_g = 0.5$.

The fermion action is given by
\begin{equation}
  S_F
  = a^4\sum_x \left [
  \bar{\psi}(x) ( i D_W + m_0 )
  \psi(x)
   + a  \csw \bar\psi(x)\frac{i}{4}\sigma_{\mu\nu}
  F_{\mu\nu}(x)\psi(x) \right ],
\end{equation}
We expect the smearing to bring the discretization errors close to the tree-level values and choose $ \csw=1$. We have performed a few short measurements of the clover coefficient and find them consistent with the tree-level value, even at small values of $\beta$.

The smeared links are calculated in three sequential stout smearing steps,
summing over the directions that are orthogonal to those in the previous steps. \cite{Durr:2010aw}.
We have chosen the smearing coefficients for each step to be
$\alpha_1 = 0.78$, $\alpha_2 = 0.61$ and $\alpha_1 = 0.35$.

We measure the running coupling using the  Shc\"odinger functional method\cite{Luscher:1992ny,Luscher:1992an}.
We use boundary conditions to the temporal direction to induce a chromoelectric background field and measure the coupling by the response to a change in the background field.
The boundary conditions are
\begin{align}
U_\mu(\bar x,t=0) = e^{-i\eta \sigma_3 a/L}, \,\,\,\,\,\, U_\mu(\bar x,t=L) = e^{-i(\pi - \eta) \sigma_3 a/L}
\end{align}
with $\sigma_3$ the third Pauli matrix and we choose $\eta=\pi/4$.
The spatial boundary conditions are periodic for the gauge field. The fermion field is set to zero at the temporal boundaries and have twisted periodic boundary conditions to the spatial directions: $ \psi(x+L\hat i) = exp (i\pi/5) \psi(x) $.

At the classical level the derivative of the action with respect to $\eta$ is
\begin{align*}
 \frac{\partial S^{cl.}}{\partial \eta} = \frac{k}{g_0^2},
\end{align*}
where $k$ is a function of $N$ and $\eta$. We define the full renormalized coupling at quantum level by 
\begin{align*}
 \ev{\frac{\partial S}{\partial \eta}} = \frac{k}{g^2},
\end{align*}

\begin{figure}
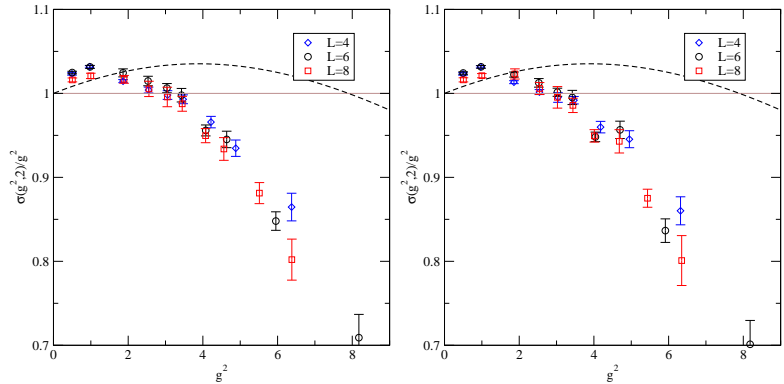

\centering
\psfig{file=su2adjsigmanorw.eps,width=0.4\textwidth}
\psfig{file=su2adjsigma.eps,width=0.4\textwidth}
\caption{
  The scaled lattice step scaling function $ \Sigma(g^2,2,L/a)/g^2 = g^2(g_0^2,2L/a)/g^2(g_0^2,L/a) $. The plot on the left shows the result using the original measurements and the plot on the right using reweighted values. The black dashed line gives the continuum 2-loop perturbative result for $ \sigma(g^2,2)/g^2 $.
}
\label{fig:latticeSigma}
\end{figure}

\begin{figure}
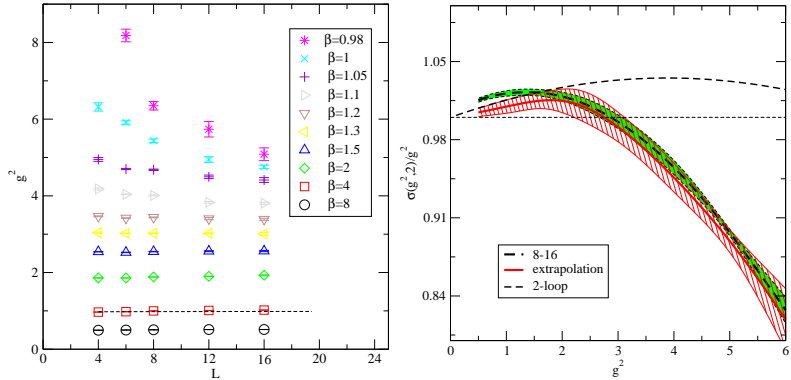

\centering
\psfig{file=su2adjrwg2.eps,width=0.4\textwidth}
\psfig{file=su2adj.eps,width=0.41\textwidth}
\caption{
  On the left, the measured values of $g^2(g_0^2,L/a)$ against
  $a/L$. 
  The black dashed line gives an example of the running in 2-loop perturbation theory at modest coupling, normalized so that it matches the measurement
  at $L/a=6$. The measurements have been reweighted to zero mass.
  On the right, the scaled step scaling function $ \sigma(g^2,2)/g^2$ using reweighted measurements. The red line with the hashed band correspond to the continuum extrapolation using lattice sizes $L=6$ and $8$ and the black line with the green band correspond to the lattice result at the largest lattice size $L=8$. The black dashed line gives the 2-loop perturbative result. 
}
\label{fig:res}
\end{figure}

The running of the coupling is quantified by the  step scaling function $\sigma$. It describes the change of the measured coupling when the linear size of the system is changed from $L$ to $sL$ keeping the bare coupling $g_0^2$ constant.
\begin{align}
 &\Sigma(u,s,L/a) = g^2(g_0^2,sL/a)|g^2(g_0^2,L/a)=u \\
 &\Sigma(u,s,L/a) = \sigma(u,s) + c(u,s) a^2 \label{eq:sigmacont}
\end{align}
We choose $s=2$ and obtain the continuum limit from measurements at $L/a=6$ and $L/a=8$.
The measured values of the coupling squared are shown in figure \ref{fig:res}.
We also show the lattice step scaling function 
$ \Sigma(g^2,2,L/a)/g^2 = g^2(g_0^2,2L/a)/g^2(g_0^2,L/a) $
on the left side in figure \ref{fig:latticeSigma}.

The Schr\"odinger functional boundary conditions reduce the amount of zero modes in the system and allow simulations at zero quark mass. Since the Wilson fermion action breaks chiral symmetry and allows additive renormalization of the quark mass, we use the PCAC relation to find the value of $\kappa$ where the renormalized mass vanishes. The quark mass $M$ is defined by
\begin{align}
aM(x_0) = \frac14 \frac{(\partial_0^* + \partial_0) f_A(x_0)}{f_P(x_0)}.
\end{align}
We define $\kappa_c$ as the value of the parameter $\kappa$ where the mass $aM(L/2)$ vanishes. To find $\kappa_c$ we measure the mass at 3 to 7 values of $\kappa$ on lattices of size $L/a=16$ and interpolate to find where the mass is zero. In practice we achieve $aM < 0.003$.

We use reweighting to correct for the residual effect of the nonzero mass. The difference can be seen in figure \ref{fig:latticeSigma}, where we show the lattice step scaling function calculated from both the original and the reweighted measurements.

The continuum limit $\sigma(g^2,2)$ is taken keeping the coupling $g^2$ equal, but the measurements have been calculated at constant $g_0^2$, and do not correspond to the same value of $g^2$ at different lattice sizes. The measurements, therefore, need to be shifted to matching values of $g^2$. The most economical and convenient way to achieve this is to interpolate the measurements at each lattice size $L/a$ by fitting to a function of $g_0^2$. This results in a value of $g^2(g_0^2,L/a)$ over a continuous range of $g_0^2$.

We use the interpolating function
\begin{align}
 \frac{1}{g^2(g_0^2,L/a)} = \frac{1}{g_0^2}\left [ \frac{ 1+\sum_{i=1}^n a_i g_0^2 }{ 1+\sum_{i=1}^m b_i g_0^2 } \right ], \label{eq:interpolating}
\end{align}
with $n=4$ and $m=2$. These values were chosen to maximize the combined $P$ value for the fit, calculated from the sum of $\chi^2$ and degrees of freedom for each fit.The interpolating functions are then used to calculate $\Sigma(u,2,L/a)$ at a continuous range of u, and the continuum limit is calculated by fitting to the quadratic function in equation \ref{eq:sigmacont}. The result is shown in figure 3.

\begin{table}[t] \centering
\tbl{The values of $\chi^2$ and degrees of freedom in the interpolation in eq. \ref{eq:interpolating}.}
{\footnotesize
\begin{tabular}{|l|l|l|l|l|l|l|}
\hline
\,
 & $L/a=4$ & $L/a=6$ & $L/a=8$ & $L/a=12$ & $L/a=16$ & combined \\
\hline
 $\chi^2$ & $32.77$ & $38.02$ & $23.71$ & $19.27$ & $11.48$ & $125.2$
 \\
 d.o.f & 4 & 5 & 5 & 5 & 5 & 24
 \\
\hline
\end{tabular}
}
\label{table:chi2}
\end{table}
 
\section{Conclusions}

Both the lattice results and the estimated continuum limit show a non-trivial infrared fixed point between $g^2=2$ and $4$. This is in agreement with previous studies.
Furthermore, a large part of the discretization errors in the interesting region seems to be removed by the clover improvement and the HEX smearing. Smearing also reduces the computation time required to generate lattices and enables simulations at larger coupling. We plan to verify this further by repeating the measurement with an increased lattice size.

\end{document}